\documentclass[journal]{IEEEtran}
\usepackage{graphicx}
\usepackage{multirow}
\ifCLASSINFOpdf

\else

\fi

\begin{document}
\title{Interference and bandwidth adjusted ETX \\
in wireless multi-hop networks}

\author{\IEEEauthorblockN{Nadeem Javaid$^{\dag}$, Ayesha Bibi$^{\ddag}$, Karim
    Djouani$^{\dag,\S}$\\\vspace{0.4cm}}
    \IEEEauthorblockA{ $^{\dag}$LISSI, Universit\'e Paris-Est Cr\'eteil (UPEC), France. \{nadeem.javaid,djouani@univ-paris12.fr\}}\\
    $^{\ddag}$ICIT, Gomal University, D.I.Khan, Pakistan. ayeshabibi@ieee.org\\
    $^{\S}$F'SATI, Pretoria, South Africa. djouanik@tut.ac.za
     }

\maketitle

\begin{abstract}
%\boldmath
In this paper, we propose a new quality link metric, interference and bandwidth adjusted ETX (IBETX) for wireless multi-hop networks. As MAC layer affects the link performance and consequently the route quality, the metric therefore, tackles the issue by achieving twofold MAC-awareness. Firstly, interference is calculated using cross-layered approach by sending probes to MAC layer. Secondly, the nominal bit rate information is provided to all nodes in the same contention domain by considering the bandwidth sharing mechanism of 802.11. Like ETX, our metric also calculates link delivery ratios that directly affect throughput and selects those routes that bypass dense regions in the network. Simulation results by NS-2 show that IBETX gives 19\char\% higher throughput than ETX and 10\char\% higher than Expected Throughput (ETP). Our metric also succeeds to reduce average end-to-end delay up to 16\char\% less than Expected Link Performance (ELP) and 24\char\% less than ETX.
\end{abstract}


\IEEEpeerreviewmaketitle

\section{Introduction}

\IEEEPARstart{W}{ireless} multi-hop networks consist of wireless nodes that are not in the transmission range of each other. So, the intermediate nodes act as routers to receive and send the routing and data packets from and to the nodes in their transmission range. In order to have appreciable performance from the underlying wireless network, the routing protocol that is responsible to operate it, plays a key role. The heart of a routing protocol is the link metric. The Minimum Hop-count is the most popular and IETF standard metric \cite{1} and is appropriately used by Wireless Ad-hoc Networks, where the objective is to find new paths as rapidly as possible in the situations where quality paths could not be found in due time due to higher rates of node mobility. Secondly, hop-count is simple to calculate and it avoids any computational burden on the routing protocol. But in the case of Wireless Mesh Networks (WMNs) mobility is not an issue; where either stationary or minimally mobile nodes interconnect and form a wireless backbone. Now, depending upon the demands of a static wireless multi-hop network; low end-to-end delay and high throughput, the routing protocol must choose a realistic routing link metric to select the quality links. Several newly proposed   metrics \cite{3}, \cite{10}, \cite{17} have succeeded to find the quality paths more efficiently than the previous ones \cite{2}, \cite{4}, \cite{5}. Since, we are dealing with the static wireless networks where all nodes are broadcasting by nature and the links do not have the same characteristics, therefore, the nodes have to compete for the transmission opportunities with their neighbors resulting in contention. Consequently, the network performance degrades mainly because of two issues; firstly, the links with lower bit rates lower the performance of faster links, secondly, the interference causes congestion and collisions that pretend the medium to be busy. Heretofore, none of the work has considered both of the phenomena simultaneously.

In this paper, we propose a new routing metric, interference and bandwidth adjusted ETX (IBETX), that selects the optimal paths in the wireless multi-hop networks. As longer paths usually achieve higher throughputs, the metric takes them into consideration while selecting the best path, unlike Expected Transmission Count (ETX) \cite{2} (and all ETX-based metrics that do not explicitly handle interference and are unable to consider longer paths). Like \cite{3}, our metric is hybrid; it is load-dependent and takes care of link quality as well. The routing layer can give appreciable performance in multi-hop networks, if it takes the relevant information from the MAC layer. For IBETX to have more accurate information, we used cross-layer to take the wireless link information from MAC layer. Then we use this information to compare the links by their transmission rates and the amount of contention they have, by measuring the interference.

The rest of paper is ordered as follows: section II states the up-to-date work on link metrics along with their deficiencies. Section III points out those shortcomings in the existing metrics that are the reasons for the motivation of this work. The section IV gives the details that how this work overcomes the shortcomings in existing metrics. Then section V details the simulation parameters chosen for this work and discusses the simulation results.
\vspace{-0.3cm}
\section{Related Work}
	
In recent years, though many quality link metrics have been proposed, still minimum hop-count is widely used by existing wireless routing protocols. Using this metric, the source node selects the least hop route to the destination node. The   metric  blindly   selects  minimum  hop   routes   without comparing  the  loss   ratios of the competing routes. This may increase the number of retransmissions causing loss in throughput and resulting degradation in the overall performance of the underlying network. To overcome this problem, ETX metric is proposed by De Couto \textit{et al.} \cite{2}. It is the expected number of (re)transmissions required to successfully transfer a packet over a link. The link interference is not taken into consideration by the ETX metric. The authors in \cite{3} proposed ELP to find optimal paths in a mesh network. To estimate link performance, ELP uses both link traffic and link quality information. It does not consider the bandwidth of the contending links. Draves \textit{et al.} \cite{4}, proposed Expected Transmission Time (ETT) that is multiplication of ETX with the link bandwidth to obtain the expected link airtime for the successful transmission of a packet. The interference is not taken into account by ETT, thus, another metric Weighted Cumulative ETT (WCETT) [4] tackles this issue along with using ETT. One of the limitations of ETX is that it may not follow the link quality variations. So, Modified ETX (mETX) and Effective Number of Transmissions (ENT) have been proposed in \cite{5} that are aware of the probe size. These metrics consider the standard deviation to project physical-layer variations along with the link-quality average values. But inter-flow interference handling mechanism is not present in WCETT. The authors in \cite{6} and \cite{7} proposed the Metric of Interference and Channel-switching (MIC). It tackles the issue of inter-flow interference and guarantees the shortest paths by calculating the interference due to the neighbors and selects the minimum-cost paths by the help of MIC virtual nodes. mETX and ENT metrics do not take into account the intra-flow interferences, therefore, Interference AWARE (iAWARE) \cite{8} estimates the average time for which the medium remains busy because of (re)transmissions from each interfering neighbor. To measure the effects of variations in the routing metrics due to continuously produced interference by neighboring nodes, this metric uses Signal to Noise Ratio (SNR) and Signal to Interference and Noise Ratio (SINR).

While counting the number of (re)transmissions required to transmit a data packet, ETX does not consider the maximum number of MAC-layer retransmissions. Therefore, Distribution Based Expected Transmission Count (DBETX) \cite{9} performing the cross-layer optimization, achieves higher network throughputs in the presence of fading while channels are continuously changing their behavior. ETT is not able to evaluate multi-channel paths precisely when the paths are long. The authors in \cite{10} proposed Exclusive Expected Transmission Time (EETT) to select multi-channel routes with the least interference when channels are distributed on a longer path to maximize the end-to-end throughput. Therefore, this metric takes into consideration the channel distribution on long paths that are critical in Large Scale Multi-radio Mesh Networks (LSMRMNs). But DBETX and EETT can not consider the longer paths due to not implementing any mechanism to calculate the interference among wireless neighboring links. ETX is not designed to consider the multi-rate links, so, Expected Data Rate (EDR) \cite{11} took Transmission Contention Degree (TCD) into account. This metric is used for making conservative estimates for paths longer than 3-4 hops by combining time-sharing effects of MAC like Medium Time Metric (MTM) \cite{12} that also minimizes the consumption time of the physical medium by avoiding longer paths.

ETX performs poor transmission bit-rate selection at the 802.11 level. Therefore, Estimated Transmission Time (EstdTT) \cite{13} assumed the size of the packet to be constant of 1500\textit{bytes} by neglecting the overhead. ETX is designed for single radio, single channel environment. For better utilizing the bandwidth in the case of multiple channels, interface switching is required and then cost of interface switching is to be considered. Multi-channel Routing Protocol (MCR) \cite{14} takes into consideration the interface switching cost and selects channel diverse routes. To improve routing path without relying the frequently broadcast route probing messages (as in original ETX), the ETX metric is combined with greedy forwarding (ETX Distance metric) in \cite{15}. But the metric makes no calculation to measure the bandwidth of the contending links and
nodes. In \cite{16}, ETX is optimized for energy-conservative networks and
named as Multicast ETX (METX). It is an energy-efficient routing metric and reduces the total transmission energy in the existence of an unreliable link layer. The bandwidth sharing of 802.11 is not taken into account by the ETX, so, Expected Throughput (ETP) is proposed by Vivek P. \textit{et al.} \cite{17}. It is a MAC-aware routing metric. This metric takes into consideration the nominal bit rates of the contending links in the neighborhood of a given link. But like ETX and ETT, it also does not consider interference. Table.1. lists the existing metrics along with the issues they have not considered.

\begin{table}[h]
\caption {SHORTCOMINGS IN ETX-BASED METRICS}
\begin {center}
\begin{tabular}{|c|c|}
\hline
\textbf{Issue(s) not considered} & \textbf{Metric} \\ \hline
\multirow{7}{*}{Inter-flow interference} & ETT \cite{4} \\
 & DBETX \cite{9}\\
 & EETT \cite{10} \\
 & EstdTT \cite{13} \\
 & MCR \cite{14} \\
 & METX \cite{16} \\
 & ETP \cite{17} \\
  \hline
\multirow{6}{*}{Bandwidth} & ELP \cite{3} \\
 & WCETT \cite{4} \\
 & MIC \cite{6},\cite{7} \\
 & iAWARE \cite{8} \\
 & EDR \cite{11} \\
 & ETX Dist \cite{15} \\ \hline
Link asymmetry & MTM \cite{12} \\ \hline
\multirow{2}{*}{Bandwidth and inter-flow interference} & ETX \cite{2} \\
 & mETX and ENT \cite{5} \\ \hline
\end{tabular}
\end{center}
\end{table}

\vspace{-0.65cm}
\section{Motivation}
This section states and discusses the weaknesses in the existing metrics that
are the reasons to propose IBETX.
The working principle behind the minimum hop-count implicitly states that whether a path works well or it doesn't work at all, it is selected among a set of available paths based on the least number of hops.
Being a non-quality link metric, it does not compare the transmission rates, packet loss ratios and interference due to neighbors on different links. Maximum network performance can be achieved by the respective routing protocol operating the underlying network. The routing protocol performs efficient routing provided that the link metric implemented with it can efficiently find quality paths. ETX augments the throughput of multi-hop paths two times as that of minimum hop-count metric by selecting the quality links \cite{2}. ETX and ETX-based metrics \cite{1} have to face many issues but we only discuss those deficiencies
that are once overcome, will improve the metric efficiency and
consequently performance of the network. These weaknesses are listed
below.

\textit{(A)} ETX sums the (re)transmission counts of all the links to find the transmission count of the entire path by assuming that all the links on that particular path contend with each other. This is true for less hop paths but is not applicable for longer paths because longer paths have more links that are not in the same contention domain \cite{17}. This spatial reuse implies that the actual transmission cost of a path is less than the sum of the transmission counts of all the links of the path. Thus, adding the ETX of all the links of a path unfairly increases the cost of longer paths due to more packet drops. In other words, ETX penalizes routes with more hops [2]. So, the metric does not consider the longer paths to select the best one. This deficiency of ETX and ETT has been depicted in figure.1. In the figure, there are three available paths from source to destination. ETX and all those ETX-based metrics that do not take inter-flow interference into account, would select one of the paths between $Path1$ and $Path3$ and would penalize $Path2$. It is obvious from the figure that $Path2$ has multiple contention domains $(CDs)$. The transmissions on a link in $CD1$ do not interfere the transmissions taking place on a link in $CD3$. As a whole, $Path2$ has interference value comparable to that of $Path1$ and $Path3$ or even less. As longer paths have higher throughput \cite{2}, \cite{12} but are ignored by ETX, so, $Path2$ is never selected for data transmissions.

\begin{figure}[h]
\begin{center}
\includegraphics[scale=0.40]{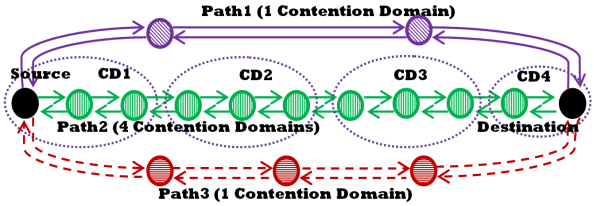}
\caption{Shortcoming of ETX and ETX-based metrics to ignore longer paths}
\end{center}
\end{figure}

\vspace{-0.3cm}
\textit{(B)} ETX, ETT and ETP do not explicitly implement any mechanism to encounter interference that usually becomes performance bottleneck in the wireless static networks.

\textit{(C)} ETX and ETT do not take any information from the MAC-layer that makes the computations more robust at the routing layer.

\textit{(D)} ETX and ELP are not capable of differentiating among the transmissions taking place on the links in the same contention domain. Being unable to calculate the bandwidth of the contending nodes, ETX and ELP do not consider the longer paths. Though, the later one takes into account the longer paths by implementing the interference but the former one still remains unable to take the longer paths into account. So, ETP tackles this issue and takes the bandwidth values of the contending links
into account. The model proposed in ETP \cite{17} considers the reduction in successful data delivery due to contention from the slow links and expects the better routes than ETX and ETT. An obvious problem of ETP, like ETX and ETT is that it does not take interference into consideration. Usman \textit{et al.} \cite{3} keeping this issue in view, proposed a new metric, ELP, that calculates interference among the wireless nodes in the same contention domain. But ELP does not provide any mechanism to take transmission rates of the contending links into account. Secondly, it increases computational burden in the algorithm by generating probes of different sizes. Thirdly, the way by which it tunes the delivery ratios (keeping $\alpha=0.75$), is useful in congested networks only, that is not always the situation.

In the next section we discuss our proposed metric that along
with measuring delivery ratios,  incorporates the two-fold MAC-layer interaction to calculate bandwidth and interference among the contending nodes.

\vspace{-0.3cm}
\section{Interference and Bandwidth Adjusted ETX (IBETX) Metric}
We understand that finding the delivery ratios is the primary quantity of interest for selecting quality links. Then comes the issue of contention due to neighbors in a wireless medium. Third most important task is to find high throughput paths that are ignored by ETX. Keeping these concerns in view, IBETX is designed as threefold metric. Firstly, it directly calculates the Expected Link Delivery (ELD), $d_{exp}$; that avoids the computational burden, as generated by ETX and bypasses the congested regions in the network like ETX. Secondly, it provides the nodes with the information of nominal bit rates and makes them able to compute Expected Link Bandwidth (ELB), $b_{exp}$, of all the wireless links in the same contention domain by cross layer approach. Thirdly, long-path penalization by ETX is encountered by calculating the interference, $I_{exp}$, named as Expected Link Interference (ELI) also by cross-layered approach. Then we define IBETX as follows:

%eq.1
\begin{eqnarray}
IBETX=\frac{d_{exp}}{b_{exp}}\times I_{exp}
\end{eqnarray}

\vspace{0.4cm}
Following sub sections give the details that how above given three mechanisms help IBETX to achieve the performance gains.

\vspace{-0.3cm}
\subsection{ELD}
This part of the metric finds the paths with the least expected number of (re)transmissions, that may be used onwards for data packet delivery. In other words, the metric estimates the number of required retransmissions calculating the delivery ratios in forward direction by $d_f$  and in reverse direction by $d_r$ of a wireless link $mn$, as given below:

%eq.2
\begin{eqnarray}
d_{exp}(mn)=d_f\times d_r
\end{eqnarray}

\vspace{0.3cm}
Besides the presence of losses, the main objective of this part is to find the paths with high throughput. To compute $d_f$  and $d_r$, each node broadcasts a probe packet (134$byte$) every second. Each probe keeps the number of probes previously received from each neighbor in the last 10$s$. Thus each node remembers the loss rates of probes on the links to all neighbors in both directions. The quantity $d_{exp}$ in addition to considering lossy links also helps to decrease the energy consumed per packet, avoiding retransmissions. It detects and suitably handles asymmetry by incorporating loss ratios in both directions. It does not route around congested links by avoiding the oscillations that cause more end-to-end delay and by selecting the routes which are either idle or they have less traffic to pass
with better delivery ratios by increasing the throughput and better utilizing the network.

This is true that $ETX =\frac{1}{d_f \times d_r}$ produces more overhead than minimum hop-count metric but this overhead is negligible, when compared to the raise in throughput. Keeping this in view, \textit{ELD}
not only achieves higher throughput values than hop-count but also over
performs ETX. Because, \textit{ELD} avoids the computational overhead generated by ETX that first takes inverse of all $d_{exp}$'s and then adds them up, whereas, \textit{ELD} only takes their sum. Our network consists of 50 nodes, where this overhead is small but in general, this overhead is directly proportional to the number of nodes or links.
This fact is depicted following in Fig. 2.

\begin{figure}[h]
\begin{center}
\includegraphics[scale=0.50]{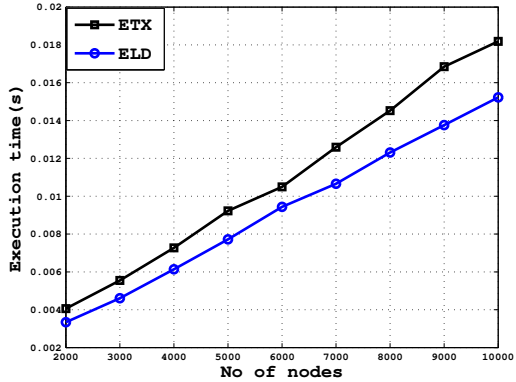}
\caption{Comparison of computational overhead generated by ETX and ELD}
\end{center}
\end{figure}

\vspace{-0.8 cm}
\subsection{ELB}
In the wireless environment, slow links lower the bandwidth of the faster ones in their neighborhood. Consequently, all contending links get the same probabilities for transmission due to underlying 802.11 Distribution Coordination Function (DCF) mechanism \cite{17}. This means that nominal bit rate information of the contending links is an important link quality factor. Suppose, we are interested to find the best path between two nodes $m$ and $n$ among a set of contending links either on a source-destination path $P$ or on a non source-destination path $NP$ but in the same contention domain. Then the expected bandwidth of the link $mn$ can be written in the following way:

%eq.3
\begin{eqnarray}
b_{exp}(mn)=\frac{1}{\sum_{i\epsilon P\cap NP}^{}\frac{1}{r_i}}
\end{eqnarray}

\vspace{0.4cm}
Here $r_i$ is the transmission rate of the $i^{th}$ link in the domain $P \cap NP$. Thus capturing the bandwidth sharing mechanism of 802.11 DCF, $b_{exp}(mn)$ considers the accurate throughput reduction of the faster links due the slower ones and predicts the better routes. Moreover, $b_{exp}(mn)$ also encounters the longer paths that are ignored by ETX and ETX-based metrics, as shown in Fig.1.

\vspace{-0.2cm}
\subsection{ELI}
	The delivery ratio $d_{exp}(mn)$ and bandwidth $b_{exp}(mn)$ calculated in the previous subsections help to directly achieve the primary objective, i.e., quality routes but they do not explicitly reveal interference of the links. Interference helps to consider the longer paths ignored by ETX and all those ETX-based metrics that do not calculate the interference among the neighbor links. To exactly measure
the congestion in the medium and collisions due to hidden nodes, interference also finds the optimal paths in the wireless network. Moreover, since the probes used to calculate $d_{exp}(mn)$ are very small in size, so, they are successfully received even in a congested network, by depicting the wrong image of link qualities. For example, if a link has only capacity to carry probe packets, it pretends the
congested link to be quality link
because of its high delivery ratios. Infect, it is not able to carry data packets \cite{3}. We, therefore, incorporate a mechanism to
calculate the interference in our metric and define \textit{ELI} that is an expected value calculated by all the nodes
on the same source-destination path.

The 802.11's basic Medium Access Control (MAC) is DCF that besides enabling the nodes to sense the link before sending data, also avoids collisions by employing the virtual carrier sensing. DCF achieves this using Request To Send (RTS) and Clear To Send (CTS) control packets that consequently set the Network Allocation Vector (NAV), i.e.,
$NAV=\tau_{RTS}+\tau_{CTS}$.
The NAV is a counter kept that is and maintained
by all nodes in the domain with an amount of time that must elapse until the
wireless medium becomes idle. Any node can not transmit until NAV becomes zero.
It stores the channel reservation information to avoid the hidden terminal problem. Using the cross-layer approach, DCF periodically probes the MAC to find the time period for which the link is busy; $\tau_{busy}$.
The interference, a node \textit{m} has to suffer, is expressed as:

%eq.4
\begin{eqnarray}
i_m=\frac{\tau_{busy}}{\tau_t}
\end{eqnarray}
\vspace{0.2cm}

Where $\tau_{busy}$ is the is the duration for which the medium remains busy; in the case of receiving packets it is $R_x$  state (or communication is going-on with other nodes) and the NAV pending.
In the interference expression for node \textit{m}, $\tau_t$ is the total window time (10$s$). If a node $n$ is at the transmitting end, its $\tau_{busy}$ is given as: $\tau_{R_x }+\tau_{T_x}+\tau_{RTS}+\tau_{CTS}$. Thus the interferences for sending node $n$ and receiving node $m$ are given as:
%\vspace{-0.2cm}

%eq.5
\begin{eqnarray}
i_m=\frac{\tau_{R_x}+\tau_{RTS}+\tau_{CTS}}{\tau_t}
\end{eqnarray}
%\vspace{0.3cm}

and

%eq.6
\begin{eqnarray}
i_n=\frac{\tau_{R_x}+\tau_{T_x}+\tau_{RTS}+\tau_{CTS}}{\tau_t}
\end{eqnarray}
%\vspace{0.3cm}

%eq.7
\begin{eqnarray}
i_{mn}=Max(i_m,i_n)
\end{eqnarray}
%\vspace{0.1cm}

The link $mn$ formed by nodes $m$ and $n$ are suffering from an interference, $i_{mn}$, that is the maximum of the interferences calculated in eq.(5) and eq.(6), is calculated by eq.(7).

The receiving node $m$ saves the information of interference computed by eq.(5) and sending node $n$ by eq.(6). Then we calculate the expected interference of the link $mn$ as:

%eq.8
\begin{eqnarray}
I_{exp}=\frac{i_{mn}}{1+i_{mn}}
\end{eqnarray}
\vspace{0.1cm}

Being shared in nature, wireless medium has a problem of interference due to contention. This causes packet loss due to collisions that consequently reduces the bandwidth of links. We, therefore, added $I_{exp}$ factor, that handles the inter-flow interference among the contending nodes. As discussed in section III, the longer paths with higher throughputs are ignored by ETX and ETX-based metrics (as shown in Fig.1), $ELI$ would not let any path (independent of number of hop-counts) to be ignored while selecting high throughput paths.

IBETX value for the end-to-end path $P$ is calculated by eq.(9), where $mn$'s are the links on $P$.

%eq.9
\begin{eqnarray}
IBETX(P)=\sum_{mn=1}^{n}IBETX(mn)
\end{eqnarray}
\vspace{0.2cm}

Then the routing metric for the best path $P_{best}$ from source to
destination is the minimum value of all available $P$'s. As given below:

%eq.10
\begin{eqnarray}
f(P_{best})=\displaystyle\min_{mn\in P_{best}}IBETX(mn)	
\end{eqnarray}
\vspace{0.2cm}

Hence, directly calculating the loss probability, expected bandwidth and expected interference based on the degree of contention present on the links, IBETX successfully finds the quality links.

\vspace{-0.3cm}
\section{Simulations}
This section provides the details concerning the simulation environment. We implement and compare the performance of our proposed metric IBETX with ELP, ETX, and ETP in NS-2.34. The window $w$ used for link probe packets is chosen to be of size 10$s$ and is named as $\tau_t$, as discussed in the last section. The wireless network consists of 50 nodes randomly placed in an area of 1000\textit{m} x 1000\textit{m}. The 20 source-destination pairs are randomly selected to generate Continuous Bit Rate (CBR) traffic with a packet of size 640$bytes$. To examine the performance of metrics under different network loads, the traffic rate is varied from 2 to 10 packets per second. For each packet rate, the simulations are run for five different topologies for 900$s$ each and then their mean is used to plot the results.

Wireless networks suffer from bandwidth and delay. Because of on-demand nature, the reactive protocols are best suited to cope with these issues for mobile scenario where change in topology is frequent. We are dealing with static networks where proactive protocols work at their best because of getting the picture of whole topology and independent of the data generation. Among the widely used proactive protocols; Destination-Sequenced Distance Vector (DSDV) \cite{18}, Fish-eye State Routing (FSR) \cite{19}, and Optimized Link State Routing (OLSR) \cite{20}, we prefer DSDV because of the following reasons:

\textit{(A)} ETX and ELP have been implemented in DSDV.

\textit{(B)} FSR and OLSR only use periodic updates that consume more bandwidth due to large size. On the other hand, along with periodic updates, DSDV also uses trigger updates. The former ones carry all the
available routing information or complete routing table, called 'full-dump'.
While the later ones merely carry 'incremental'. An incremental is an information changed since last 'full-dump'and is well
fitted in a Network Protocol Data Unit (NPDU).
The trigger updates help DSDV to reduce routing overhead that raises throughput.

\textit{(C)} Like \cite{2} and \cite{3}, our implementation further
enhances DSDV to never send 'full-dump' with trigger
updates, called 'no-dumps', rather the full-dumps are sent merely at the 'full-dump' periods.

\textit{(D)} FSR's 'graded frequency' mechanism along with the 'fish-eye' technique works better in dense networks and OLSR's Multi-point Relay (MPR) is suitable in static and dense networks. But for our simulations the case is contrary, as our network consists of 50 nodes.

\textit{(E)} FSR and OLSR when receive any data packet, they immediately send it at the already calculated route. But on receiving a data packet, DSDV waits for a duration of WST (Weighted Settling Time) during which, if it finds some better route (provided by trigger or periodic update), it sends data on that route. This mechanism works well for quality link metrics.

\begin{figure}[h]
\begin{center}
\includegraphics[scale=0.50]{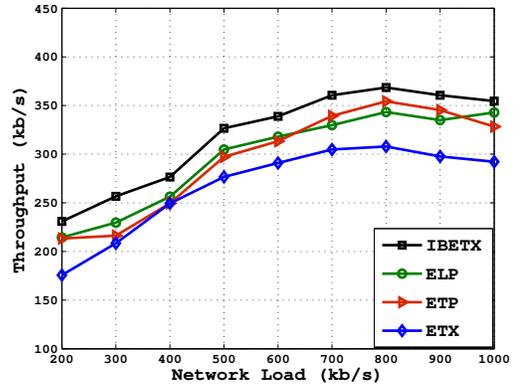}
\caption{Comparison of avg throughput achieved by DSDV with four metrics}
\end{center}
\end{figure}
\vspace{-0.3cm}

In the absence of any mobility, IBETX achieves higher throughputs than the other three metrics chosen for the comparison; ETX, ETP and ELP, as shown in Fig.3. This performance is achieved due to implementing the multiple performance criteria in IBETX. In static wireless multi-hop networks, all nodes prefer the shorter paths and as a result the underlying network experiences congestion. Thus, to accurately measure the link quality is more important in static networks. Measuring the probability of success for data packet delivery using probes is more useful strategy when compared only with shortest path. So, the part $ELD$ nicely performs the job and achieves higher throughput than ETX by avoiding the computational overhead of taking inverse of the probability of success $(d_f\times d_r)$ for all links, as shown in Fig.2.

But as the network under analysis is wireless by nature, the links with lower bit rate degrade the performance of the faster links. Therefore, taking the bandwidth of all links in the same contention domain into account gives more accurate information about the link status as compared to simple considering probability of success.
In eq.(3), $b_{exp}$ tackles this issue by implementing the bandwidth sharing mechanism of 802.11 DCF and considers the throughput reduction of faster links due to contention of slower links. Consequently, $b_{exp}$ predicts quality links and helps IBETX to achieve increased throughput as compared to ELP, ETP and ETX, as obvious from Fig.3. The metric achieves
19\char\% more throughput than ETX, 10\char\% more than both ETP and ELP.
As, probes are smaller in size as compared to data packets, so, the idea
of measuring the link quality by calculating probability of success along with nominal bit rate information does not suffice. Therefore, our metric incorporates the interference part that rightly predicts the medium congestion and collisions due to hidden nodes that increase the end-to-end delay. To accurately estimate the medium occupation, using cross-layered approach, \textit{ELI} periodically probes the MAC-layer 100 times per second. In MAC broadcast probes, all nodes in the network piggyback their interferences for the last $\tau_t seconds$ (10$s$), hence, \textit{ELI} avoids extra routing overhead. Moreover, since, IBETX
can consider longer paths due to \textit{ELB} and
\textit{ELI}, it increases throughput and reduces E2ED.
So, our metric reduces end-to-end delay up to 15\char\% lower than ETP, 16\char\% lower than ELP and 24\char\% lower than ETX; because only ELP directly implements interference
but has some performance leaks, as not measuring bandwidth of the contending
nodes and varying probe size, etc. Comparison of end-to-end delay produced
by all of the competing four metrics is depicted in Fig.4.

 \begin{figure}[h]
\begin{center}
\includegraphics[scale=0.50]{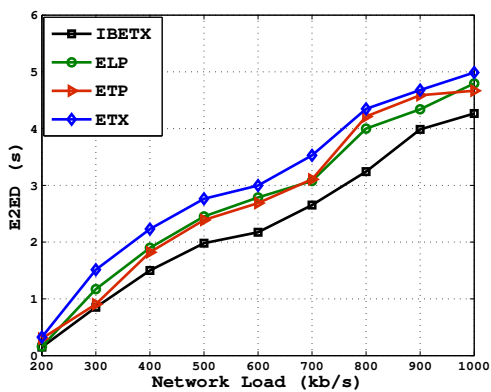}
\caption{Average End-to-end delay produced by DSDV with four metrics}
\end{center}
\end{figure}
\vspace{-0.3cm}

\section{Conclusion and Future Work}
In this work, we proposed a new quality link metric for wireless multi-hop networks. We have overcome the performance leaks in ETX due to its unawareness from the MAC layer. Using cross-layer approach, we provided our metric with the MAC layer information. $ELD$ found the high throughput paths more efficiently than ETX and ELP by avoiding the overhead due to computational complexities in both. $ELB$ found the quality links from all active links in the same contention domain. \textit{ELI} part along with \textit{ELB} removed the deficiency in ETX and ETX based metrics
to ignore the longer paths while selecting quality links, though the longer paths usually give higher throughputs. In future, we are interested to enhance the functionality of IBETX to work in multi-channel environment.
Moreover, because of the computational overhead reduction by \textit{ELD},
IBETX can achieve even higher throughput values, if it is implemented
with OLSR in network with more population of nodes.

\ifCLASSOPTIONcaptionsoff
  \newpage
\fi


\begin{thebibliography}{00}

\bibitem{1}Javaid, N. \textit{et al.}, "Performance Study of ETX Based Wireless Routing Metrics," 2nd International Conference on Computer, Control and Communication, pp. 1-7, 2009.
\bibitem{2}De Couto \textit{et al.}, "A High-Throughput Path Metric for Multi-Hop Wireless Routing," in ACM Mobicom, 2003.
\bibitem{3}Usman A. \textit{et al.}, "An Interference and Link-Quality Aware Routing Metric for Wireless Mesh Networks,". IEEE 68th Vehicular Technology Conference, 2008.
\bibitem{4}Richard D. \textit{et al.}, "Routing in Multi- Radio, Multi-Hop Wireless Mesh Networks," in ACM Mobicom, 2004.
\bibitem{5}Koksal C. \textit{et al.},"Quality-aware routing metrics for time-varying wireless mesh networks", IEEE Journal on Selected Areas in Communications, vol. 24, no. 11, pp. 1984-1994, Nov. 2006.
\bibitem{6}Yang. Y. \textit{et al.},"Interference-aware Load Balancing for Multi-hop Wireless Networks", Tech. Rep. UIUCDCS-R- 2005-2526, Department of Computer Science, University of Illinois at Urbana-Champaign, 2005.
\bibitem{7}Yang. Y. \textit{et al.}, "Designing Routing Metrics for Mesh Networks", WiMesh, 2005.
\bibitem{8}Subramanian. A. \textit{et al.}, "Interference Aware Routing in Multi-Radio Wireless Mesh Networks", Technical Report, Computer Science Department, Stony Brook University, 2007.
\bibitem{9}DO Daniel. \textit{et al.}, "An Enhanced Routing Metric for Fading Wireless Channels". IEEE Wireless Communications and Networking Conference-IEEE WCNC.
\bibitem{10}Jiang. W. \textit{et al.}, "Optimizing Routing Metrics for Large-Scale Multi-Radio Mesh Networks", pages (1550-1553), WiCom, 2007.
\bibitem{11}J. Park \textit{et al.}, "Expected Data Rate: An Accurate High- Throughput Path Metric For Multi-Hop Wireless Routing", in proc. of IEEE Communications Society Conference on Sensor and Ad Hoc Communications and Networks, SECON 2005.
\bibitem{12}B. Awerbuch \textit{et al.}, "High throughput route selection in multi-rate ad hoc wireless networks", Springer Mobile Networks and Applications, vol. 11, no. 2, pp. 253-266, April 2006.
\bibitem{13}Aguayo. D. \textit{et al.}, "A high throughput routing protocol for 802.11 mesh networks (draft)", Internet Article: http://pdos.csail.mit.edu/ rtm/srcrr-draft.pdf, 2006.
\bibitem{14}P. Kyasanur \textit{et al.}, "Routing and Link-layer Protocols for Multi-Channel Multi-Interface Ad hoc Wireless Networks", Mobile Computing and Communications Review, 10(1): pp. 3 1-43, Jan. 2006.
\bibitem{15}Wang. C. \textit{et al.}, "Optimizing End to End Routing Performance in Wireless Sensor Networks", Springer-Verlag DCOSS 2007, LNCS 4549, pp. 36-49, 2007.
\bibitem{16}Q. Dong \textit{et al.}, "Minimum energy reliable paths using unreliable wireless links", In Proc. of ACM MobiHoc, 2005.
\bibitem{17}Vivek P. \textit{et al.}, "MAC-Aware Routing in Wireless Mesh Networks ".In IEEE/IFIP WONS 2007, January 2007. Obergurgl, Austria.
\bibitem{18}Perkins. C. \textit{et al.}, "Highly dynamic   destination sequenced distance-vector routing (DSDV) for mobile computers," in ACM SIGCOMM, 1994.
\bibitem{19}Pei, G. \textit{et al.}, "Fisheye state routing: A routing scheme for ad hoc wireless networks", IEEE International Conference on Communications, p70-74, 2000.
\bibitem{20}Clausen, T. \textit{et al.}, "The optimized link state routing protocol, evaluation through experiments and simulation", IEEE Symposium on" Wireless Personal Mobile Communications, 2001.


\end{thebibliography}
\end{document}